# Synthesis of InAs/CdSe/ZnSe core/shell1/shell2 structures with bright and stable near-infrared fluorescence

*Assaf Aharoni*[†,‡], *Taleb Mokari*[†,‡], *Inna Popov*[‡] *and Uri Banin*[†,‡,*]

Institute of Chemistry and the Center for Nanoscience and Nanotechnology the Hebrew University of Jerusalem, Jerusalem 91904, Israel

*To whom correspondence should be addressed: banin@chem.ch.huji.ac.il

[†] Institute of Chemistry.

[‡] The Center for Nanoscience and Nanotechnology.

## Abstract

A complex InAs/CdSe/ZnSe Core/Shell1/Shell2 (CSS) structure is synthesized, where the intermediate CdSe buffer layer decreases strain between the InAs core and the ZnSe outer shell. This structure leads to significantly improved fluorescence quantum yield as compared to previously prepared core/shell structures and enables growth of much thicker shells. The shell growth is done using a layer-by-layer method in which the shell cation and anion precursors are added sequentially allowing for excellent control and a good size distribution is maintained throughout the entire growth process. The CSS structure is characterized using transmission electron microscopy, as well as by X-ray diffraction and X-ray-photoelectron spectroscopy which provide evidence for shell growth. The quantum yield for CSS with small InAs cores reaches over 70% - exceptional photoluminescence intensity for III-V semiconductor nanocrystals. In larger InAs cores there is a systematic decrease in the quantum yield, with a yield of ~40% for intermediate size cores down to a few percent in large cores. The CSS structures also exhibit very good photostability, vastly improved over those of organically coated cores, and transformation into water environment via ligand exchange is performed without significant decrease of the quantum yield. These new InAs/CdSe/ZnSe CSS nanocrystals are therefore promising near-IR chromophores for biological fluorescence tagging and optoelectronic devices.

## Introduction

The need for nanocrystals with bright and stable fluorescence for various applications covering biology[1-4] to electrooptics[5-8] is increasingly growing. This is particularly true for III-V semiconductor nanocrystals that can cover the technologically important visible to near infrared (NIR) spectral ranges[9-16]. Moreover III-V semiconductor nanocrystals are not as developed as II-VI semiconductor nanocrystals and their optical capabilities have not yet reached their full potential.

The main strategy to increase photoluminescence (PL) quantum yield (QY) and stability of nanocrystals is to grow a passivating shell on the cores surface[11-20]. This removes surface defects acting as traps for the carriers and therefore reduces the probability for the undesired processes of emission quenching via non-radiative decay. Moreover, the passivating shell protects the core and reduces surface degradation. Two main factors are considered while choosing the semiconductor material for the passivating shell. The first is the lattice mismatch between the core and shell materials. A large lattice mismatch will cause strain at the core-shell interface that can lead to creation of defect sites acting as trap sites for the charge carriers. The second factor is the band offsets between core and shell regions that should be sufficiently high so that carriers are confined into the core region and kept separated from the surface where defects can lead to the undesired non-radiative relaxation processes. This latter effect is particularly critical in III-V semiconductors that typically are characterized by small effective masses for the charge carriers requiring a large potential barrier for their confinement. In the present case of InAs cores, the electron effective mass, $m_e^*$, is only $0.024m_e$ ($m_e$ is the free electron mass) and the breadth of the electron wavefunction is therefore significant.

Earlier works on core/shell nanocrystals resulted in QY (quantum yield) values of up to 90% for II-VI/II-VI core/shell structures[18,19] and up to 20% for III-V/II-VI core/shell structures[11,12,21]. For the III-V structures there is still significant room for improvement in the QY values but even for the II-VI structures showing high QY, the shell thickness corresponding to these maximal quantum yields is small, typically only of about 2 ML's (monolayers). This limitation is likely due to traps created by structure imperfections formed in the growth process. A thick shell is important for the stability of the nanocrystals, especially for applications in which they are exposed to tough processes, e.g. upon dissolving nanocrystals in water for biological tagging applications.

A solution to this problem was given in the work of Li et al.[22], in which a layer-by-layer growth method was used, and termed Successive ion layer adsorption and reaction (SILAR). A layer-by-layer growth was previously also used to create CdS/HgS/CdS quantum-dot – quantum-well structures[23-25]. In this method the cation and anion shell precursors are added sequentially into the reaction vessel, i.e. only when the reaction of one kind of atom is completed its counter ion is introduced. Using this method 5 ML's of CdS shell were grown on CdSe cores without quenching the PL QY. Another solution to the problem of increased stress with shell thickness is to grow a hetero-shell structure in which a buffer layer is used to decrease stress in the shell[26]. This was recently implemented by Talapin et al.[27] in growing a ZnSe buffer layer between a CdSe core and a ZnS shell. Using ZnSe, with a lattice spacing that is intermediate between that of CdSe and ZnS, they managed to lower shell strain and could grow 7 ML of shell without a significant decrease in the PL QY. Another buffer layer approach was successfully applied, also recently, by Xie et al.[28] in which the buffer layer is an alloy of the semiconductor type constituting the core and the outer shell.

In this work we have successfully synthesized a complex core/shell1/shell2 structure of InAs/CdSe/ZnSe that has CdSe as a buffer layer (Figure 1), by implementing the SILAR growth method for III-V semiconductor cores. The InAs core and ZnSe outer shell materials have a large lattice mismatch of ~7%. In order to reduce the undesirable strain effects between the InAs core and the ZnSe outer shell a CdSe buffer layer is grown in between them as the lattice mismatch between InAs and CdSe is actually nearly 0%. The InAs-ZnSe band offsets of 1.3 eV for the conduction band and 0.99 eV for the valence band[29] enables sufficient confinement of electron and hole wave functions in the InAs core, even for the extremely light electron of InAs. This new core/shell/shell (CSS) structure yields exceptional fluorescence quantum yield and stability covering the entire NIR spectral range from 800 nm to over 1.6 microns. The fluorescence QY is still very high even upon ligand exchange and transformation to water environment. This clearly demonstrates the strength of III-V semiconductor core-heteroshell nanocrystals. This is due to the ultimate flexibility afforded by the hetero-shell approach to address both the requirement of lattice matching between core and shell materials while at the same time achieving proper band offsets between the core and shell regions.

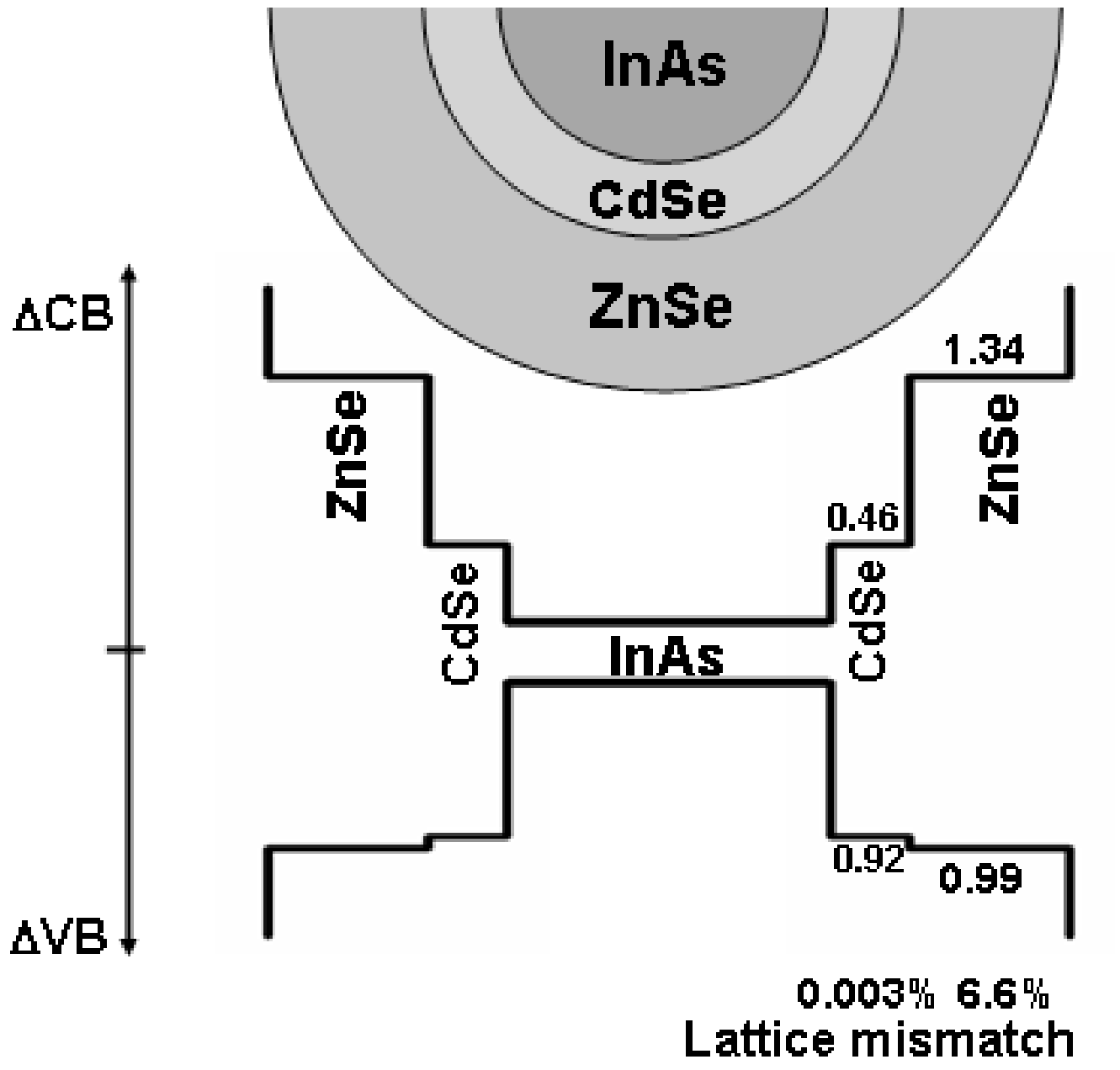

**Figure 1.** Diagram of potential structure for InAs/CdSe/ZnSe CSS structure. Band offset (in eV) and lattice mismatch (in %) between InAs and the two II-VI semiconductors used as the shell material are indicated. CB = conduction band; VB = valence band.

## Experimental Details

**Chemicals for shell growth and water solubility.** Cadmium oxide (CdO, 99.99%, powder), selenium (Se, 99.999%, powder), dimethylzinc ($Zn(CH_3)_2$, 2M in Toluene), Sulfur (99.98%, powder), 1-octadecene (ODE, 90%), oleic acid (OA, 90%), octadecylamine (ODA, 97%) and anhydrous chloroform (99+%) were purchased from Aldrich. Mercaptopropionic acid (MPA, 99%) was purchased from Fluka. All the materials were used without further purification. ODE was pumped under vacuum at 250 $^{o}$C for 2 hours before use.

**Synthesis of InAs/CdSe/ZnSe CSS (Core/Shell1/Shell2).** Details of InAs nanocrystal synthesis serving as cores is reported elsewhere[10]. In a typical synthesis of the CSS, $1x10^{-7}$ mol of InAs cores of a desired size dissolved in 700mg toluene were put in a flask with 5g of ODE and 1.5g of ODA. The amount of InAs cores was deduced from the weight. The preparation of the reaction flask is performed in a glove box where all the stock solutions are kept under strict inert air and water free conditions. The reaction flask is then transferred to a Schlenk line and pumped under vacuum until no bubbling appears (for about 20 min). Next the flask is heated gradually to 100 $^{o}$C still under vacuum to remove residues of volatile solvents. The vacuum is switched to Ar flow and the temperature is raised to 260 $^{o}$C. At around 200 $^{o}$C the first aliquot is taken (shown as 'layer 0' in the experimental results below). The first stock insertion containing Cd precursors is introduced at 260 $^{o}$C. Following, the Se is introduced. This could be continued sequentially to complete the desired thickness of shell1. Afterwards, precursors of shell2 are added sequentially, Zn precursor is first inserted followed by Se precursor and so on. The insertion continues in a variable manner between cations and anions. The time interval between each insertion is about 15min. The calculations of the amounts needed for each insertion is described below. Between stock insertions aliquots are taken to monitor the reaction.

**Cd stock solution.** 0.04 M Cd in ODE was prepared by heating CdO (154 mg) and Oleic acid (2.71g) in ODE (27 ml) at 250 $^{o}$C under Ar until a colorless solution is obtained (about 30 min).

**Se stock solution.** 0.04 M Se in ODE was prepared by heating Se (95 mg) in ODE (30 ml) at 200 $^{o}$C under Ar until all the Se powder has dissolved and a yellow clear solution is obtained (about 2 hours).

**Zn stock solution.** 0.04 M Zn in ODE was prepared by mixing $Zn(CH_3)_2$, 2M in Toluene (0.20 ml) with ODE (9.8 ml) in a glovebox.

**Calculations of stock solution amounts.** The amount of cation or anion stock solutions added for each layer is equal to the number of cations or anions present at the surface of all the particles in solution. This number is calculated taking into account particle size, amount, bond lengths and the densities of the semiconductors being grown. For the purposes of this calculation, every full anion-cation layer is considered to be 0.35 nm in thickness, the (111) lattice spacing in zinc-blende InAs. For example, when growing a shell on $1.0x10^{-7}$ moles of InAs QD's with an average diameter of 3.8 nm we used $1.8x10^{-5}$ moles of Cd and Se for the first layer (first and second insertions, named as 1/2 and 1 layers, respectively), $2.6x10^{-5}$ moles of Zn and Se for the second layer (third and fourth insertions, named as 3/2 and 2 layer, respectively) and $4.1x10^{-5}$ moles of Zn and Se for the third layer (fifth and sixth insertions, named as 5/2 and 3 layers, respectively). These amounts are the basis for the synthesis and were optimized for the highest QY as details below.

**Hydrophilic ligand exchange.** The CSS's were transferred into water by using the phase transfer method[30]. In the first stage, the ODA-coated nanocrystals were dissolved in 1 mL of anhydrous chloroform (OD = 1). Then 1 mL of methanol solution containing 0.5 mL MPA (mercaptopropionic acid) was added under vigorous stirring to the chloroform solution. After 30 seconds of stirring, 1 mL of distilled water and 0.1 ml of KOH (pH = 1) was added to the Methanol/Chloroform solution. The CSS's were completely transferred into the water phase after 20 min. The hydrophilic phase containing the nanocrystals was separated. To remove excess MPA from the solution the particles were separated by centrifugation. The precipitate was dissolved in pH 10 buffer water and yielded a clear solution.

**Optical characterization.** Absorption measurements were taken with a Jasco V-570 spectrophotometer and the photoluminescence (PL) measurement with an Ocean – optics NIR-512 spectrophotometer using 634 nm HeNe laser excitation. QY measurements were done using a Labsphere integrating sphere based on the method described in Ref. 31. We used 1064 nm CW-Nd-YAG laser for excitation and the sample PL and laser emission were both detected using an InGaAs detector. In this method, the excitation laser output from the integrating sphere is measured along with the PL and corrected for the detection system response. The QY is calculated from a ratio of the number of photons emitted extracted from the PL feature and the number of photons absorbed extracted from the laser feature with and without sample. A third measurement in which the laser does not pass directly through the sample, but the sample is placed inside the sphere, is used to deduct and correct for signal obtained from diffused laser light striking the sample over its entire area.

**Structural characterization.** Transmission electron microscopy (TEM) images were taken using a Phillips Tecnai 12 microscope operated at 100 kV. High-resolution TEM (HRTEM) images were taken on a Tecnai F20 $G^2$ operated at 200 kV. Samples for TEM were prepared by depositing a drop of sample-toluene solution onto 400 mesh copper grids covered with a thin amorphous carbon film, followed by washing with methanol to remove the excess organic residue. XRD (X-ray diffraction) measurements were performed on a Philips PW1830/40 X-ray diffractometer operated at 40 kV and 30 mA with Cu $K_{\alpha}$ radiation. Samples were deposited as a thin layer on a low-background scattering quartz substrate. X-ray photoelectron spectroscopy (XPS) was carried out with a Kratos Analytical

Axis Ulta XPS instrument. Data were obtained with Al Kα radiation (1486.6 eV). Survey spectra were collected with 160 eV pass energy detection. Survey time was 120s for each measurement. The measurements were performed on nanocrystal films of monolayer thickness, linked by hexane dithiols to a Au coated substrate[12,32].

## Results and Discussion

**Synthesis methodology.** Optimization of synthesis conditions was determined by the photoluminescence (PL) characteristics i.e. the emission intensity and width. Addition of stock solution to the growth solution was done when there was no further increase in PL intensity. Typically, a ~10 min interval was used between the first and second insertion and up to 20 min between the thirteenth and fourteenth insertion. For shell growth on smaller (< 3 nm) and bigger dots (> 4 nm) it was necessary to optimize the stock amount used for each half shell growth, using the calculated amount as a reference point. The factors obtained are 1 (4 nm dots) to 0.8 (7 nm dots) for big dots and 1 (4nm dots) to 0.5 (2 nm dots) for small dots. The optimization is necessary due to errors in the determination of the initial core amount and the loose definition of a shell thickness as described in the experimental section.

To monitor the progress of the shell growth we measured the PL during the reaction. Figure 2a shows the PL evolution as a function of insertion number of shell stock solution grown on a 3.8 nm (diameter) InAs core sample. The spectra shown were taken after each stock addition using the time intervals mentioned above.

The first shell type grown is CdSe and the external shell was of ZnSe. Already upon addition of the Cd stock, which was the first injected species, we observe a significant increase in the PL (Fig. 2b). The increase in PL intensity persists as the growth of the layers continues, leading to an overall enhancement by a factor of 47 (Fig. 2c). This increase is a result of the high potential barrier imposed by the shell which enables efficient confinement of electron and hole wave functions in the InAs core and thus diminishes their presence in the vicinity of dark traps located mostly at the nanocrystal surface area.

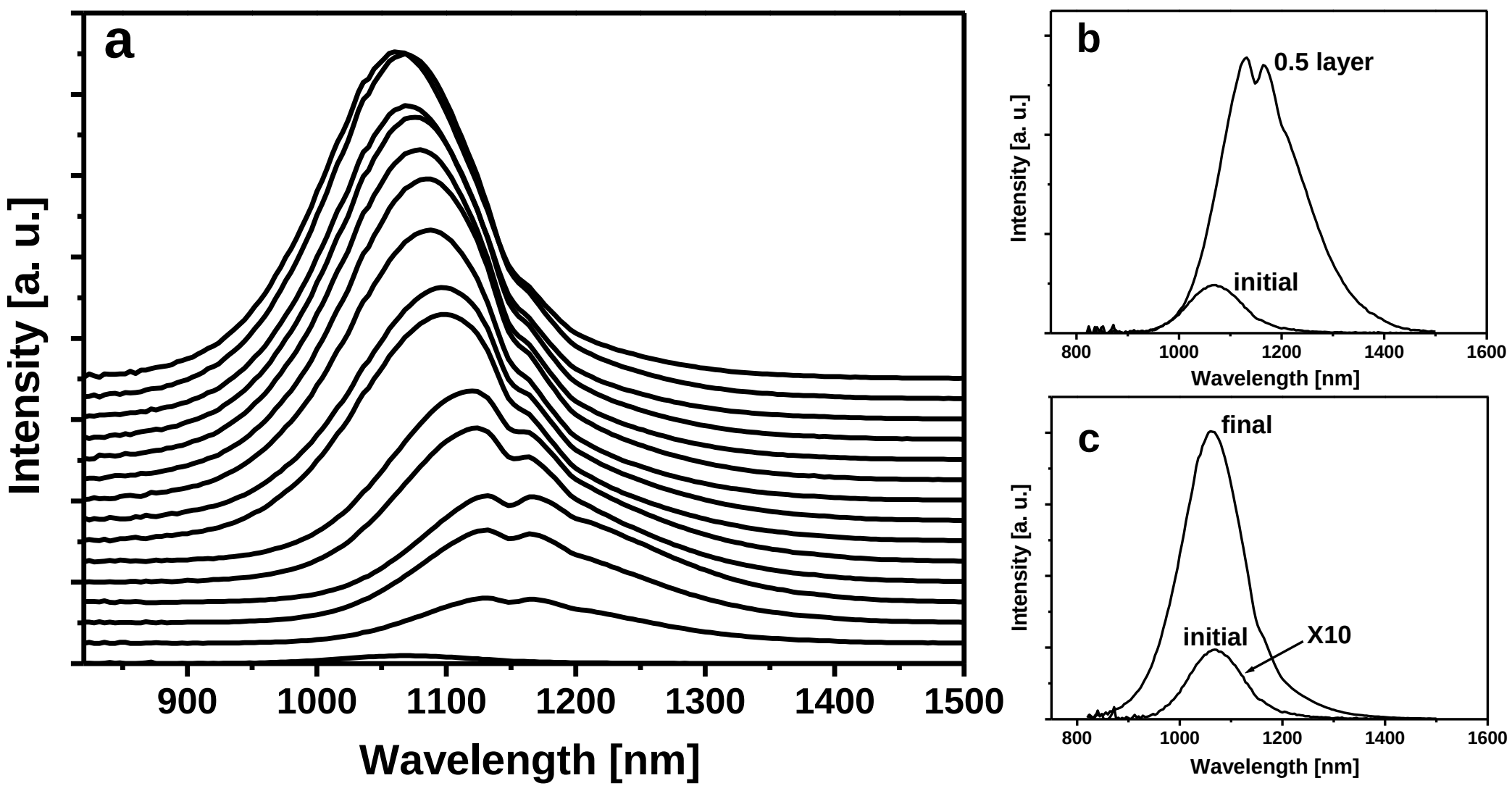

**Figure 2.** Evolution of the PL spectra of 3.8 nm InAs cores as a function of stock insertions. **a**. From bottom up, PL spectra taken after each insertion (0, 1, 2, … , 14) of stock. **b**. PL spectrum before and after the first insertion of Cd atoms. **c**. Initial PL spectrum, multiplied by 10, versus the final PL spectrum.

Two interesting phenomena are observed in addition to the significant increase in PL intensity. The first is the red shift of the PL after the growth of the first CdSe layer. This is due to the low potential barrier imposed by the CdSe shell on the InAs core, in particular for the conduction band (CB, see Fig. 1), allowing tunneling of the electron wavefunction to the surrounding shell effectively meaning the box is larger. After the first insertion of Cd precursors there is a large red shift followed by a significant increase of the PL intensity (figure 2b). The second phenomenon is the return of the PL wavelength to its original position after the growth of the ZnSe shell (figure 2c). Another notable comparison for the shell growth process is shown in Figure 3 presenting the initial and final absorption and PL of the cores and CSS's discussed above. The spectral features are hardly changed indicating the monodisprse sample size distribution is maintained throughout the complex shell growth process.

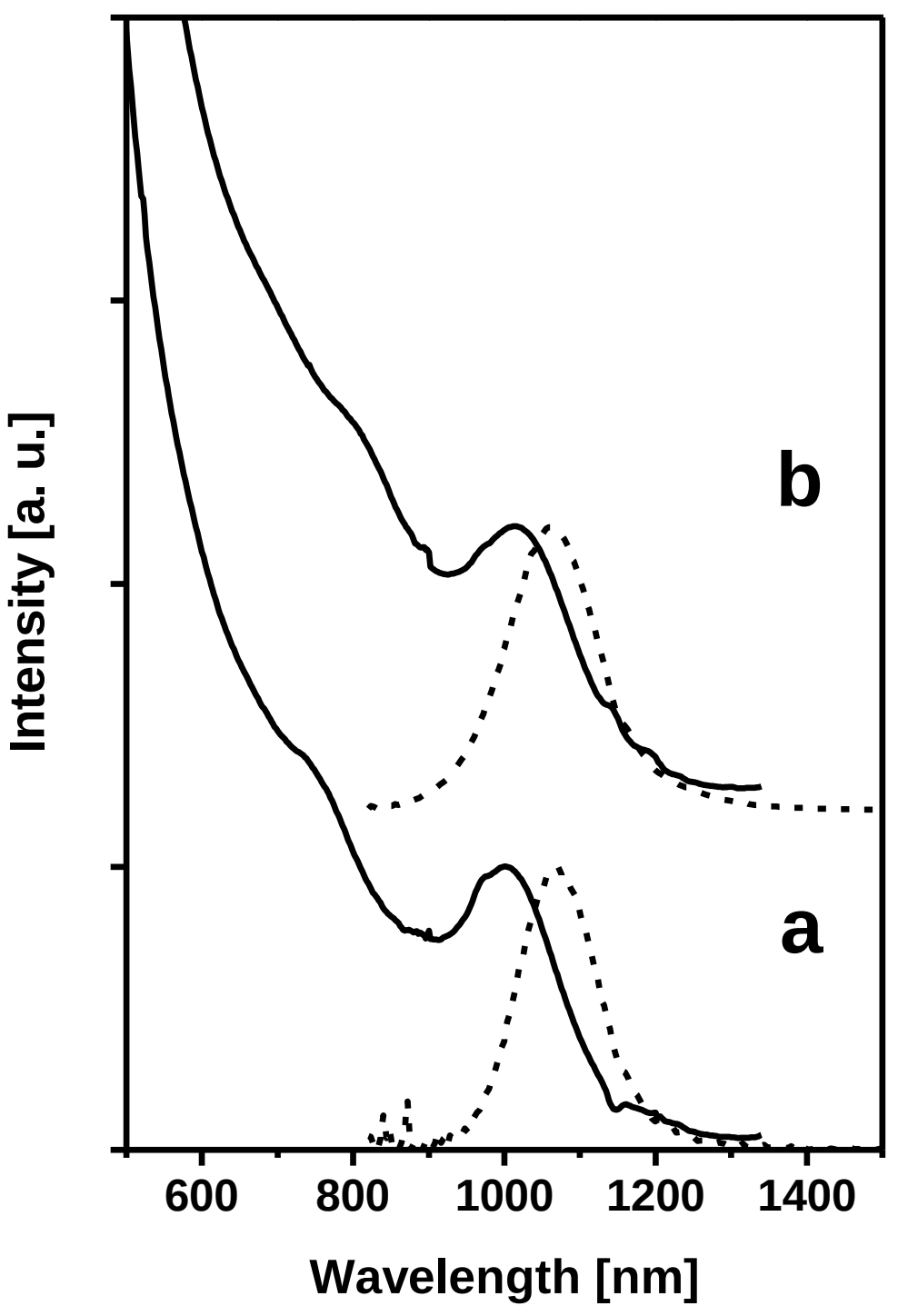

**Figure 3.** Absorption (solid lines) and PL (dashed lines) spectra of the same cores as in figure 1. **a.** Before shell growth **b.** After growth of 1.8 nm of CdSe/ZnSe shells. The spectral features are hardly changed indicating the sample homogeneity is maintained during shell growth.

Figure 4 shows the quantitative results of the PL evolution shown in figure 1. The PL Quantum Yield increases from 1% before shell growth to 50% at the end of the growth process (figure 4a, circles). This final QY value is significantly larger then the values that we could achieve using a single shell approach in earlier work (maximum of ~20%)[11,12]. Additionally, the value is achieved for a thick shell that provides significant protection for the cores.

The PL maximum red shifts from 1070 nm before shell growth to 1130 nm after the first CdSe layer and as the ZnSe shell growth progresses, the PL blue shifts back close to its original position (figure 4, top). The FWHM of the PL during shell growth is also plotted, providing indication for the size-distribution. The small change between the initial and final FWHM provides further evidence for maintaining the good original distribution as already indicated by the absorption and will be directly seen in the TEM images below.

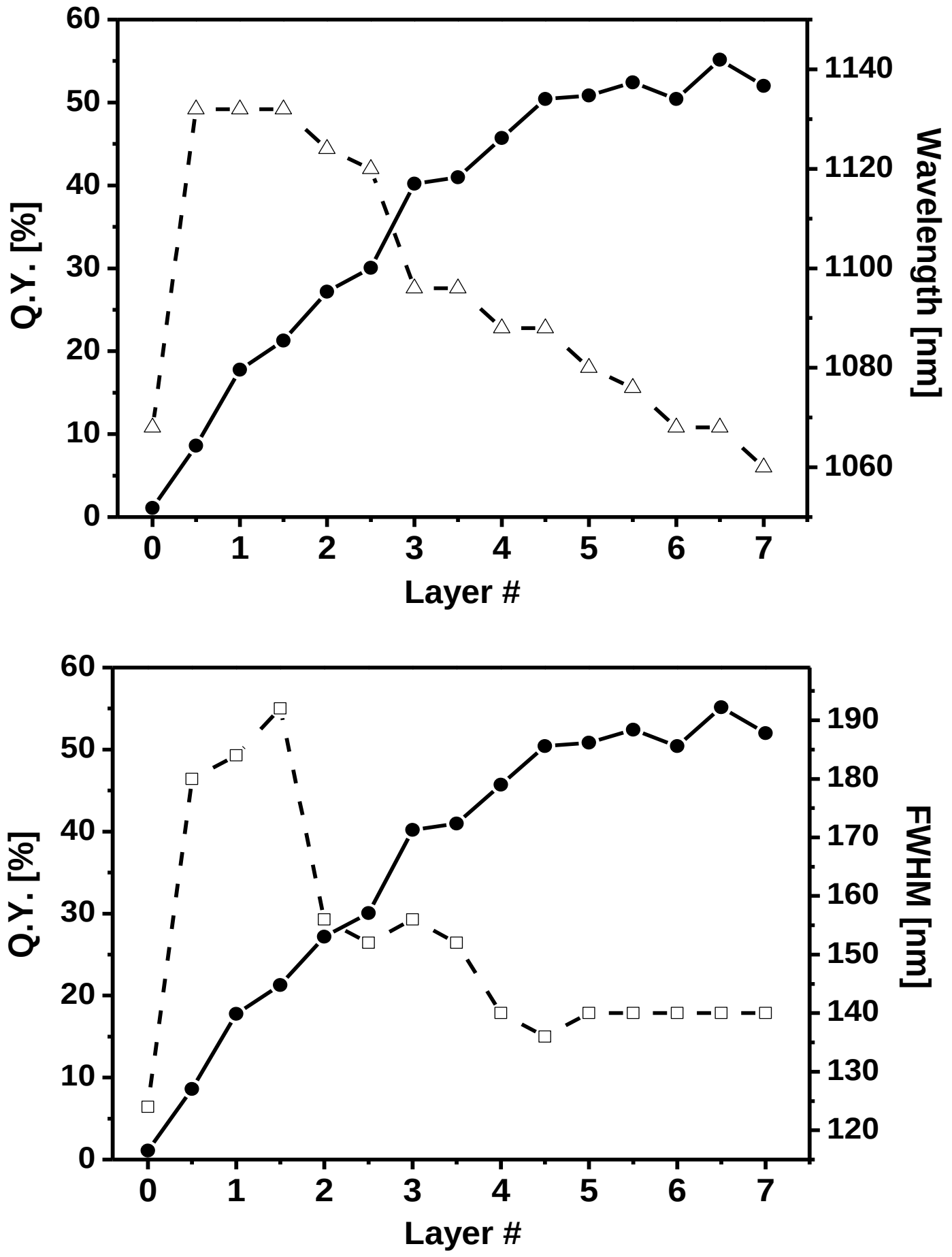

**Figure 4. Top:** Quantitative representation of the PL evolution with layer growth for CSS nanocrystals. Circles connected by lines: Quantum Yield of 3.8 nm InAs dots as a function of insertion number. The first layer is the CdSe buffer layer and the following layers are ZnSe. Open triangles connected by dashed lines: shift of PL wavelength as a function of insertion number. The growth of the first CdSe layer is followed by a red shift of the PL which blue shifts back to its initial position after the growth of the following ZnSe layers. **Bottom**: Circles connected by lines - QY same as top. Open squares connected by dashed lines: FWHM of the PL feature versus insertion number. An increase in FWHM is observed after the first insertions and is compensated as the ZnSe shell growth continues.

In order to optimize reaction conditions and procedure we have examined several factors including reaction temperature and the number of stock insertions of each semiconductor grown. When considering temperature for shell growth one has to take into account two main effects. The first is the crystalline quality of the shell grown. Using lower temperature will result in the growth of a structurally imperfect shell with traps quenching the PL. On the other hand performing the reaction at high temperature can result in nucleation of the shell precursors. The temperature used in our reaction (260 $^{o}$C) gave significantly better results in terms of QY then those obtained for growth at lower temperatures of 240 $^{o}$C (not shown) and we had no indications of a nucleation process competing with shell growth, which is a benefit of the approach of sequential anion-cation additions. i.e. in this approach each species is reacted and consumed before the counterpart is added and therefore the probability for nucleation of nanocrystals of the shell material is significantly reduced.

Extrinsic effects, i.e. the environment surrounding the particles, on the PL characteristics provides information of the effectiveness of the shell barrier. In the case of organically passivated core nanocrystals, the chemical separation process from the growth solution containing excess ligands can result in a significant loss of Q.Y. due to the removal of the passivating organic ligands. In the present case, not only is the emission not weakened during separation, in fact we have seen a small increase of 8% in PL intensity in an as prepared CSS's sample after separation via methanol precipitation and re-dissolution in toluene. This clearly demonstrates the effectiveness of the double-shell barrier confining the exciton and keeping it away from the surface area. The small increase in QY may be due to impurities probably coming from the ODE or ODA used in the reaction that are cleaned out by the separation process.

In order to verify the essential role of the combination of the two semiconductors in the core/shell1/shell2 structure to achieve superior emission properties, we have grown each of them individually on InAs cores using the SILAR growth approach. Figure 5 shows the results obtained by growing a shell containing CdSe or ZnSe exclusively.

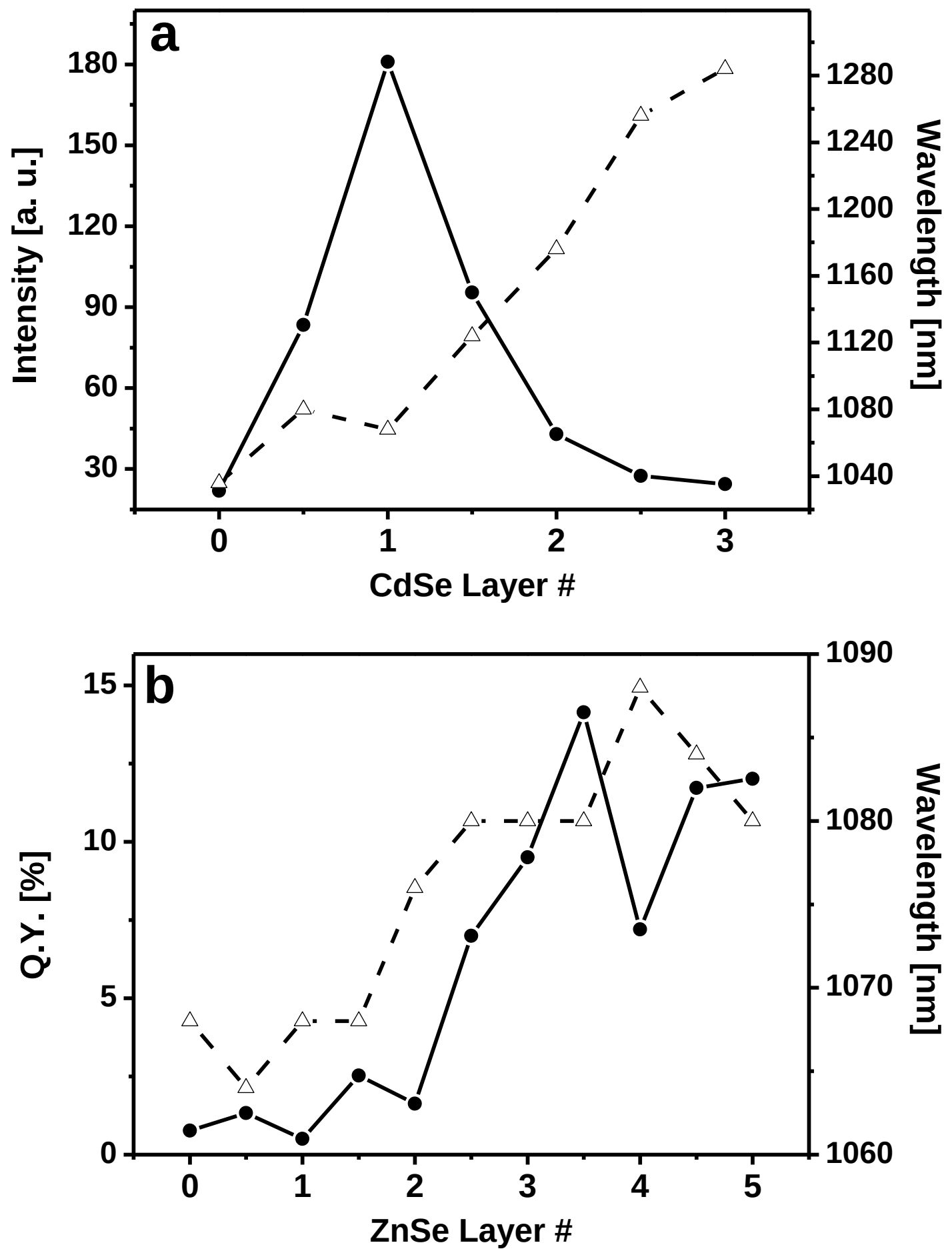

**Figure 5.** PL intensity and wavelength of InAs cores as a function of CdSe (a) and ZnSe (b) shell growth using SILAR. **a.** Circles connected by lines: PL Intensity of 3.5 nm InAs dots as a function of CdSe shell layers. Shell growth was done at 240°C. Open triangles connected by dashed lines: shift of PL max wavelength as a function of shell layers. The intensity reaches its maximal value at approximately 1ML after which it decays substantially. **b.** Circles connected by lines: Quantum Yield of 3.8 nm InAs dots as a function of ZnSe shell layers. A CdSe buffer layer was not used. Open triangles connected by dashed lines: shift of PL wavelength as a function of shell layers.

For CdSe shell the intensity reaches its maximum enhancement already at 1ML (Figure 5a) after which it decays substantially in agreement with our previous results where the CdSe shell precursors were added together[11,12] and not in the layer-by-layer method applied here. The PL wavelength red shifts as the shell grows as a result of low potential barrier of the CdSe semiconductor, leading to reduced quantum confinement. The result of growing an exclusive ZnSe shell via a layer-by-layer method is shown in figure 5b. We used the same cores used in the synthesis shown in figure 2 and similar reaction conditions. As expected the intensity increases but the final QY achieved is only 12%, close to values also achieved in the older approach of shell growth using injection of stock containing both Zn and Se precursors[12]. This demonstrates the necessity of the CdSe buffer layer in obtaining high yield NC's, most likely because of its role in relaxing the surface tension.

To further demonstrate the usefulness of the layer by layer growth in comparison to the older shell growth technique in which the cation and anion precursors are inserted simultaneously to the growth solution[11,12], we grew a CdSe/ZnSe shell on InAs cores using the same precursor amounts for each insertion and the same reaction condition except for the simultaneous insertion of shell precursors. Using this approach the PL intensity decays rapidly after the third insertion (equal to 6 insertions using the layer by layer approach) and decays completely to its original intensity after the 4'th insertion (equal to 8 insertions using layer by layer). In addition, there is an increase of 80% to the FWHM of the PL feature. In contrary to this, CSS's synthesized using the layer by layer technique show increasing PL intensity up to the 14'th insertion and the size distribution is nearly maintained as seen by absorption, FWHM of the PL and direct TEM study (see below).

By tuning the core nanocrystal size, spectral tuning for the emission can be achieved throughout the NIR spectral range via the effect of quantum confinement in the cores. Figure 6 shows the spectral coverage provided by four CSS's synthesized using cores emitting at 880, 1060, 1170 and 1420 nm.

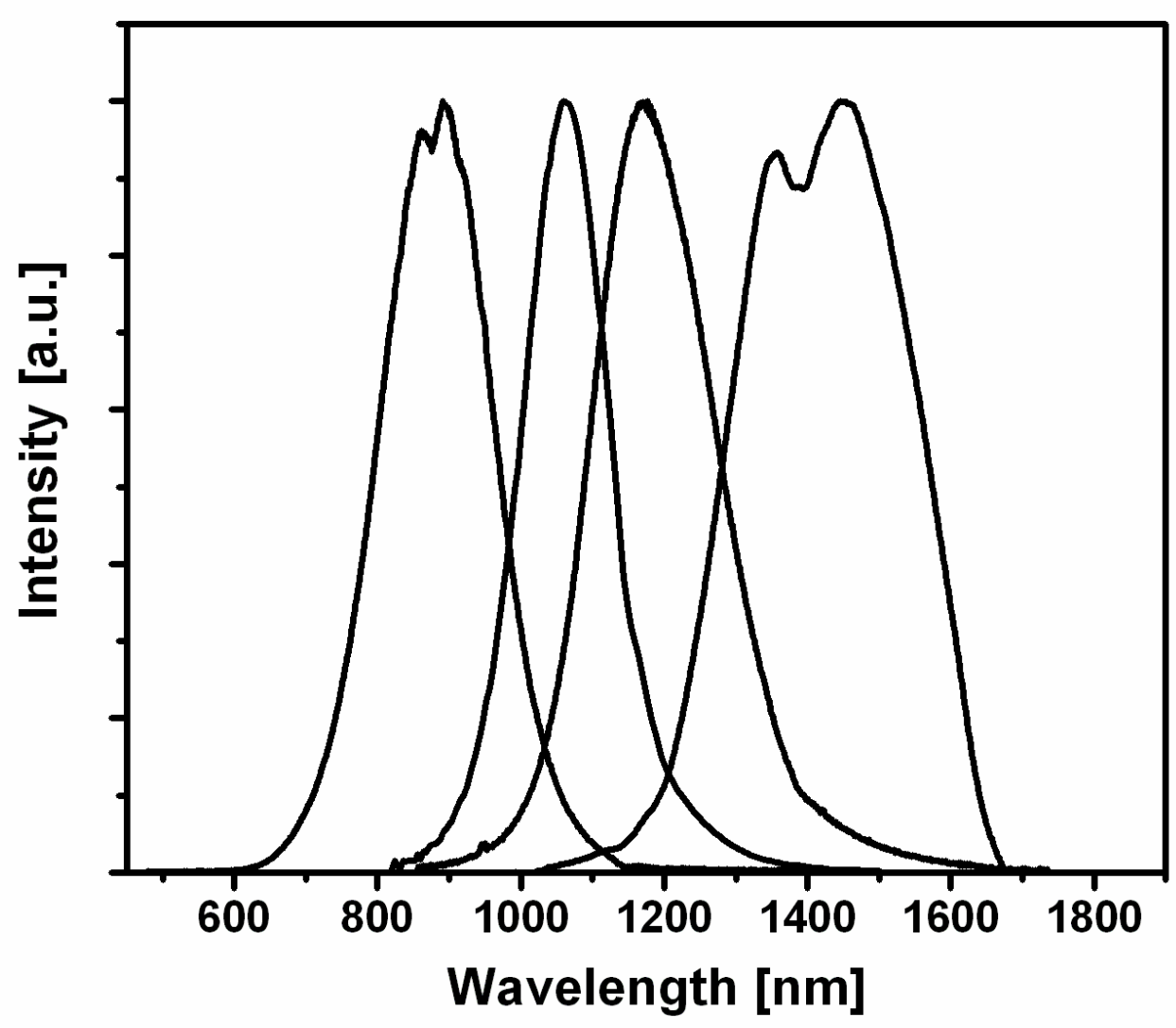

**Figure 6.** Normalized PL emission of four CSS's covering the NIR region. The PL peak position is determined by the size of the cores being used. Small features in the PL of the small and large samples are due to grating and detector features that could not be completely corrected.

Table 1 summarizes the results obtained for three sizes of CSS's ranging from 1.9 nm to 6.3 nm. The nominal shell thickness is the theoretical thickness the shell should have if the shell growth was complete assuming an increase of 0.35 nm for each layer in the [111] growth direction. The actual shell thickness received is closer to nominal for the small particles then the big ones. While the actual thickness is 80% from nominal for the 1.9 nm core particles, it is only 31% for the 6.3 nm particles. This can be explained by the high surface activity of small particles and as a consequence a higher chemical yield of shell growth.

Even more significant is that the initial QY of the cores we used is also size dependent and changes from ~ 5% for the smallest cores to lower then the detection limit for the biggest core. In correspondence to this, the final QY is higher for the smaller CSS's. There is a systematic decrease in QY going from small CSS's (more then 70%) to large (less then 2%) that is correlated with the starting QY of the cores used. There is a substantial increase in the QY for large cores attained by the SILAR growth method, but still not enough to equal the intensities reached by the small CSS's in correlation with the initial small value of the core QY. For example, the PL intensity of a core emitting at 880 nm is enhanced by a factor of 20 while in a large core emitting at 1420nm it is enhanced by a factor of 70. The lower yield at the larger diameter nanocrystals is believed to be due to an intrinsic effect of reduction in electron-hole overlap due to the very different effective masses of the electron and hole in InAs. This leads to a reduced electron-hole overlap in larger cores, and was noted also to lead to reduced QY with increased length in InAs rods[33,34]. This effect is under further study with more detailed spectroscopic investigation.

| Core size [nm] | Nominal Shell Thickness [nm] | Final Size [nm] | PL max [nm] | Initial QY [%] | Final QY [%] |
|---|---|---|---|---|---|
| **1.9** | **3.5** | **4.7** | **885** | **4.1** | **70** |
| **3.8** | **4.9** | **5.6** | **1065** | **1.1** | **52** |
| **6.3** | **4.9** | **7.8** | **1425** | **<<1** | **2.5** |

**Table 1.** Summary of PL maximum wavelength, shell thickness and QY obtained for various sizes of CSS's.

**Chemical and structural characterization.** Figure 7 shows TEM and HRTEM images of InAs cores before (frame a) and after (frame b) shell growth and histograms of size distributions. The size distributions were determined by measuring more than 300 particles of each sample. The core size is 3.8 nm and it grows to 5.6 nm with the addition of the shell. The standard deviation of the size distribution of both the cores and the CSS's are 14%. The unchanged size distribution is a measure of the excellent kinetic control of the shell growth process in this layer-by-layer approach.

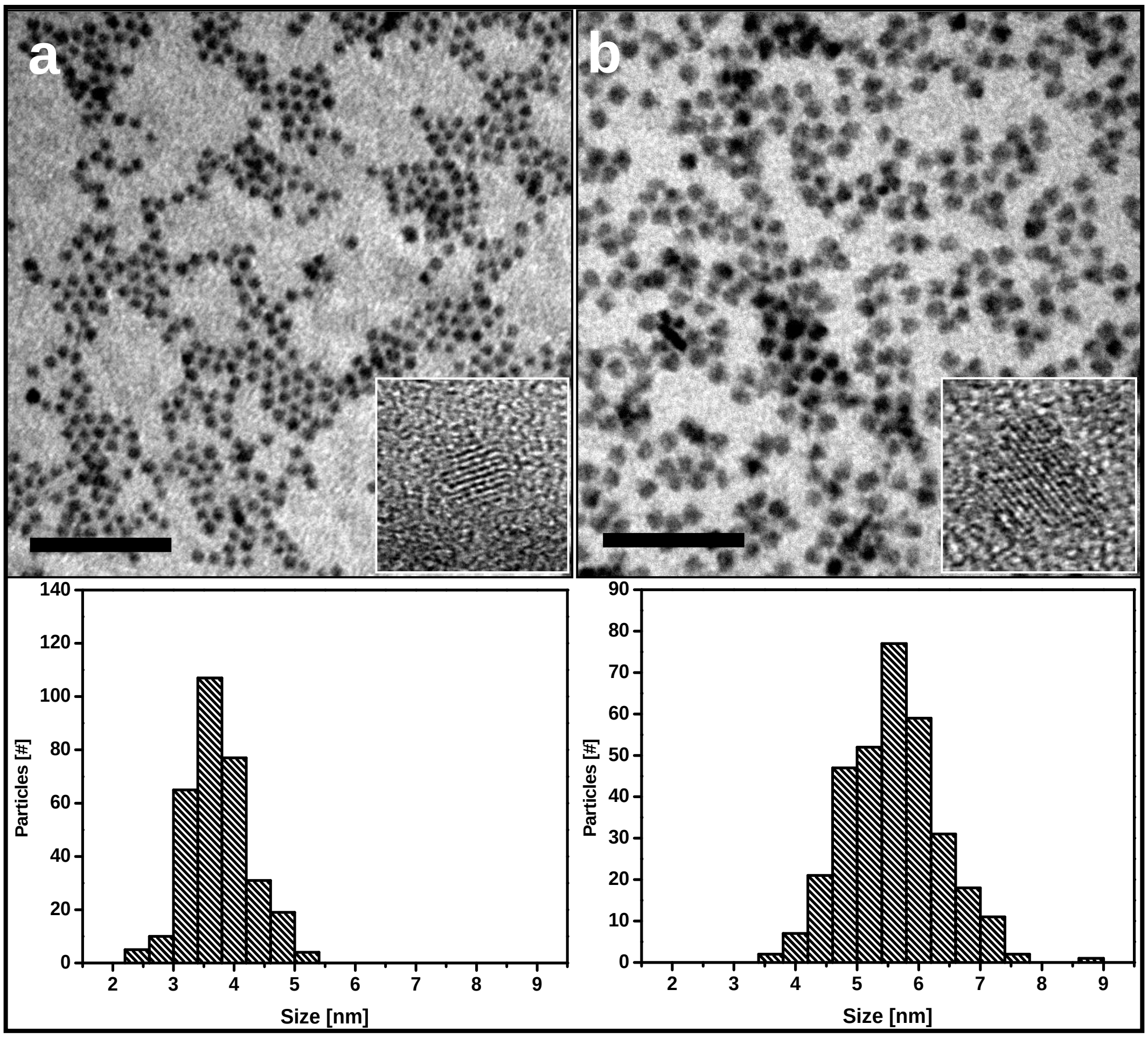

**Figure 7. a**. TEM and HRTEM (insets) of 3.8 nm InAs cores and the corresponding size distribution on the bottom. **b**. TEM and HRTEM of 5.6 nm InAs/CdSe/ZnSe CSS QD's synthesized using the InAs cores in **a** and the corresponding size distribution on the bottom. Scale bars are 50 nm. HRTEM square insets are 10x10 nm. Average sizes and standard deviation are 3.8 ± 0.5 nm for cores and 5.6 ± 0.8 nm for CSS.

To determine the crystalinity of the CSS we measured their XRD spectra. Figure 8 shows the evolution of the XRD spectra of an InAs/CdSe/ZnSe CSS sample as a function of shell layers and type. The XRD spectrum of the initial InAs cores (bottom spectrum) matches that of the bulk zinc-blende InAs. As an intermediate CSS we grew an InAs/CdSe/ZnSe CSS sample with approximately 0.5 nm of CdSe/ZnSe shell (middle spectrum). The XRD spectrum of the intermediate is shifted to larger angles in regard to the InAs core spectrum. In addition to the shift to larger angles we observe peaks matching the $In_2O_3$ BCC lattice. This lattice may be formed at the InAs core surface at high temperature before the shell growth process begins. The upper spectrum is of the same CSS sample shown in figure 7 (0.9 nm of CdSe/ZnSe shell). The CdSe amount in samples **b** and **c** is the same. Spectrum **c** continues the shift to larger angles towards the position of the bulk peaks of ZnSe zinc-blende indicating the growth of a ZnSe shell.

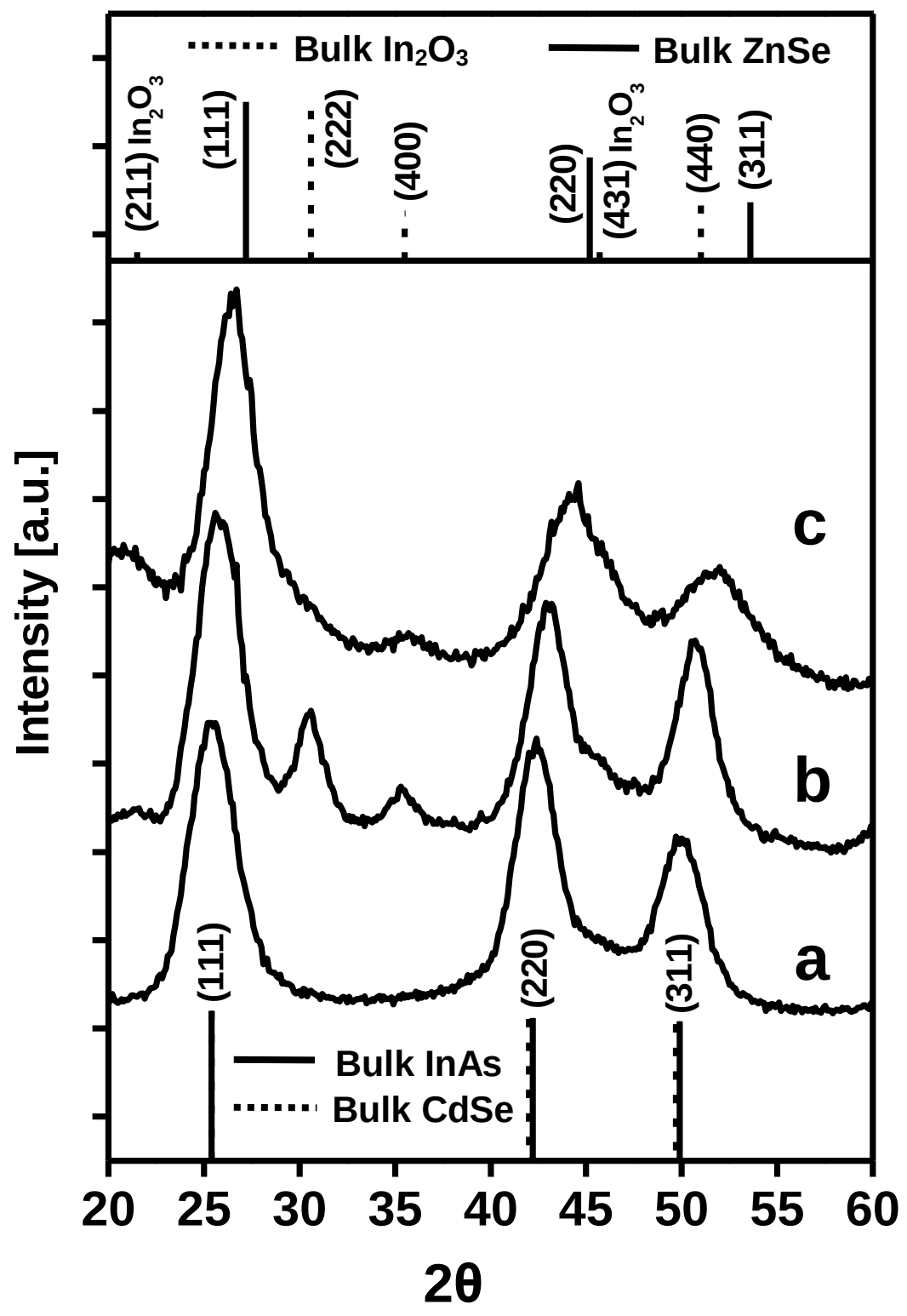

**Figure 8.** **a**. X-ray powder diffraction spectra of 3.8 nm InAs cores. **b**. InAs cores covered with approximately 0.5 nm of CdSe/ZnSe shell. **c**. InAs cores covered with 0.9 nm of CdSe/ZnSe shell. The lines represent bulk XRD patterns of zinc-blende InAs, CdSe, ZnSe and BCC $In_2O_3$. The XRD patterns shift from **a** to **c** towards bigger angles indicating the ZnSe shell growth.

Further indication of shell growth is given by X-ray photoelectron spectroscopy (XPS) measurements probing the surface of the CSS's. Figure 9 shows XPS survey measurements of InAs cores (Bottom spectrum) and InAs/CdSe/ZnSe CSS's (upper spectrum). Being a surface sensitive technique, the XPS signal depends strongly upon the distance of the material's constituent atoms from the surface of the nanocrystals. In the InAs cores spectrum the signal of various atom levels of In and As are clearly seen as well as of Au, the substrate on which the cores are deposited. However, after shell growth the In and As atoms signal, being now screened by the shell, decrease significantly or even disappear. Instead Zn and Se peaks from the outer shell are strongly dominant in the spectrum.

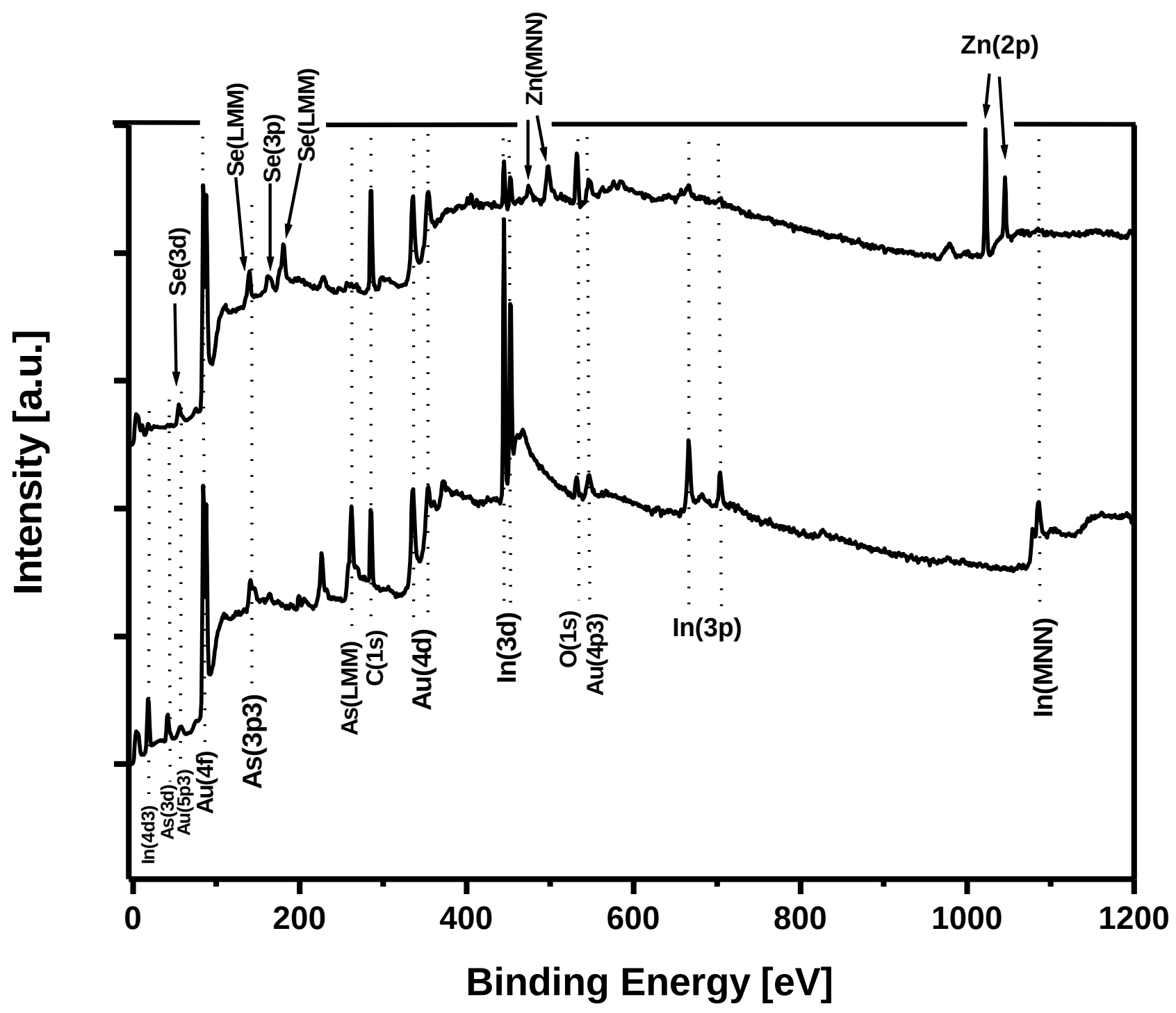

**Figure 9.** XPS survey of 3.8 nm InAs cores (bottom) and 5.6 nm InAs/CdSe/ZnSe CSS's QD's synthesized using the same cores (top). The assignment of the peaks is indicated. The peaks associated with In and As are missing or reduced in the spectrum of the CSS QD's while new peaks of Zn and Se appear, indicating growth of shell material.

**Core/shell/shell stability upon irradiation and hydrophilic ligand-exchange:** The significantly improved characteristics of the CSS particles are exemplified clearly not only by the significant enhancement in QY but also in the improved stability. Figure 10 shows the comparison of stability as reflected by relative QY measurements of the original InAs cores and CSS's made using them (same cores and CSS's as in previous figures). For this experiment the solutions containing the nanocrystals in ambient atmosphere were irradiated by a laser at 473 nm and intensity of 30 mW while stirring the solutions to ensure homogenous exposure to the laser. The emission from the solutions was measured over a period of six hours. Cores show a rapid decay and in fact completely degrade after about 1 hour and precipitated out of the solution. Under the same conditions CSS nanocrystals show a significantly better performance. After an initial decrease of the QY, the emission stabilizes at about 60% of the initial QY, still with an absolute QY of 31%. This is maintained for 5 hours with only slight slow decay and the particles remain very stable in solution.

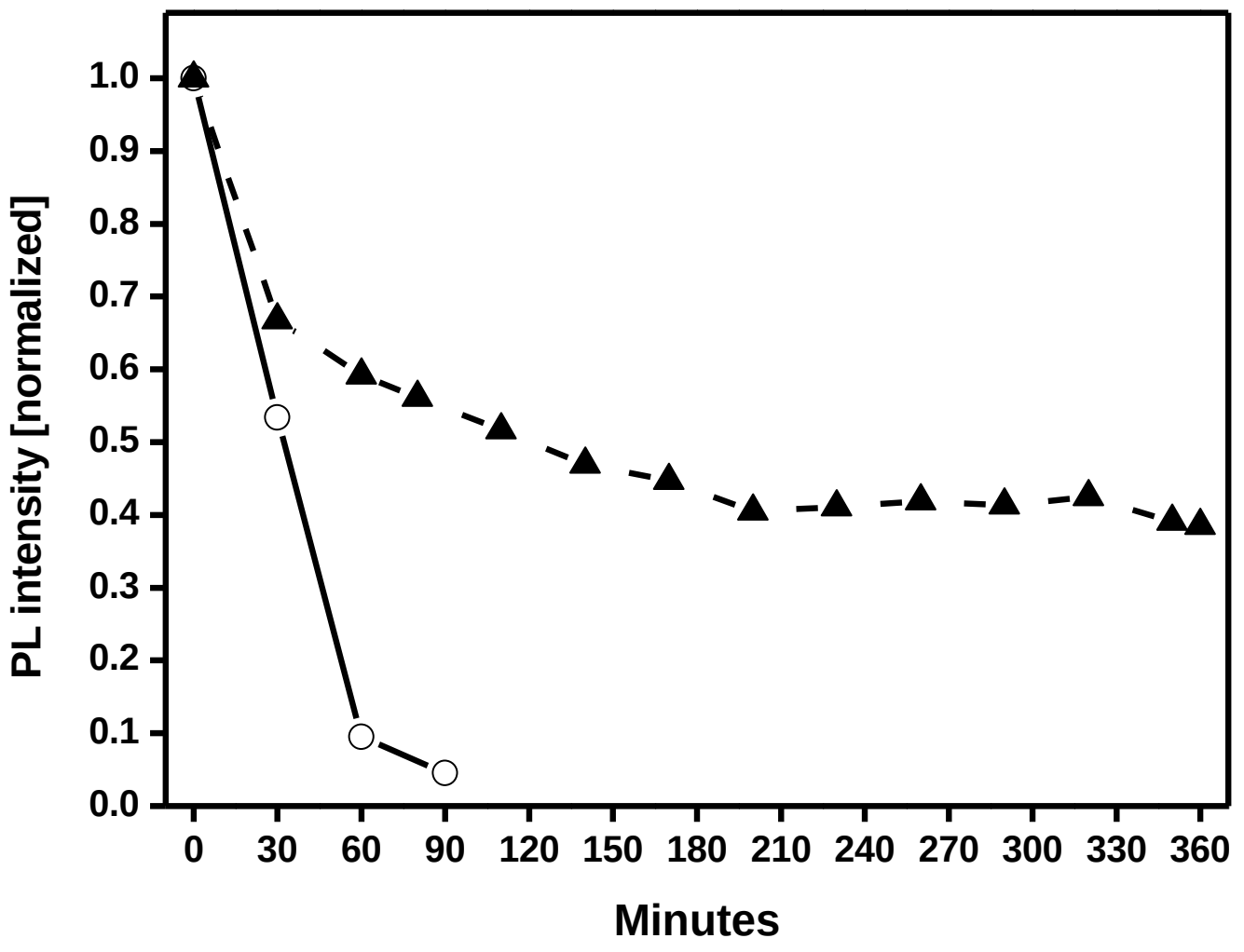

**Figure 10.** Photochemical stability of InAs cores (Open circles connected by line) and CSS (Triangles connected by dashed line). The cores and CSS used are the same shown in the other figures. The samples were dissolved in toluene and irradiated with 30 mW of 473 nm light under ambient atmosphere.

In order to examine the possibility of using these particles as taggants in biological systems we have transferred them to water. III-V semiconductor nanocrystals are of interest for biological applications because of significantly decreased autofluorescence in the NIR. Moreove, NIR emission is potentially useful for in-vivo imaging on account of its high permeability through biological tissues[4]. The phase transfer method[30] enabled us to transfer as prepared CSS's into water with very high yield. The mercapto part of the MPA connects well to NC's surface and the hydrophobic carboxylic end enables the dissolution in water. During the process a 47% decrease in CSS's quantum efficiency occurred still leaving the particles with nearly 40% QY (figure 11). This shows the high potential of using the new CSS nanocrystals in bio-imaging applications in the NIR range.

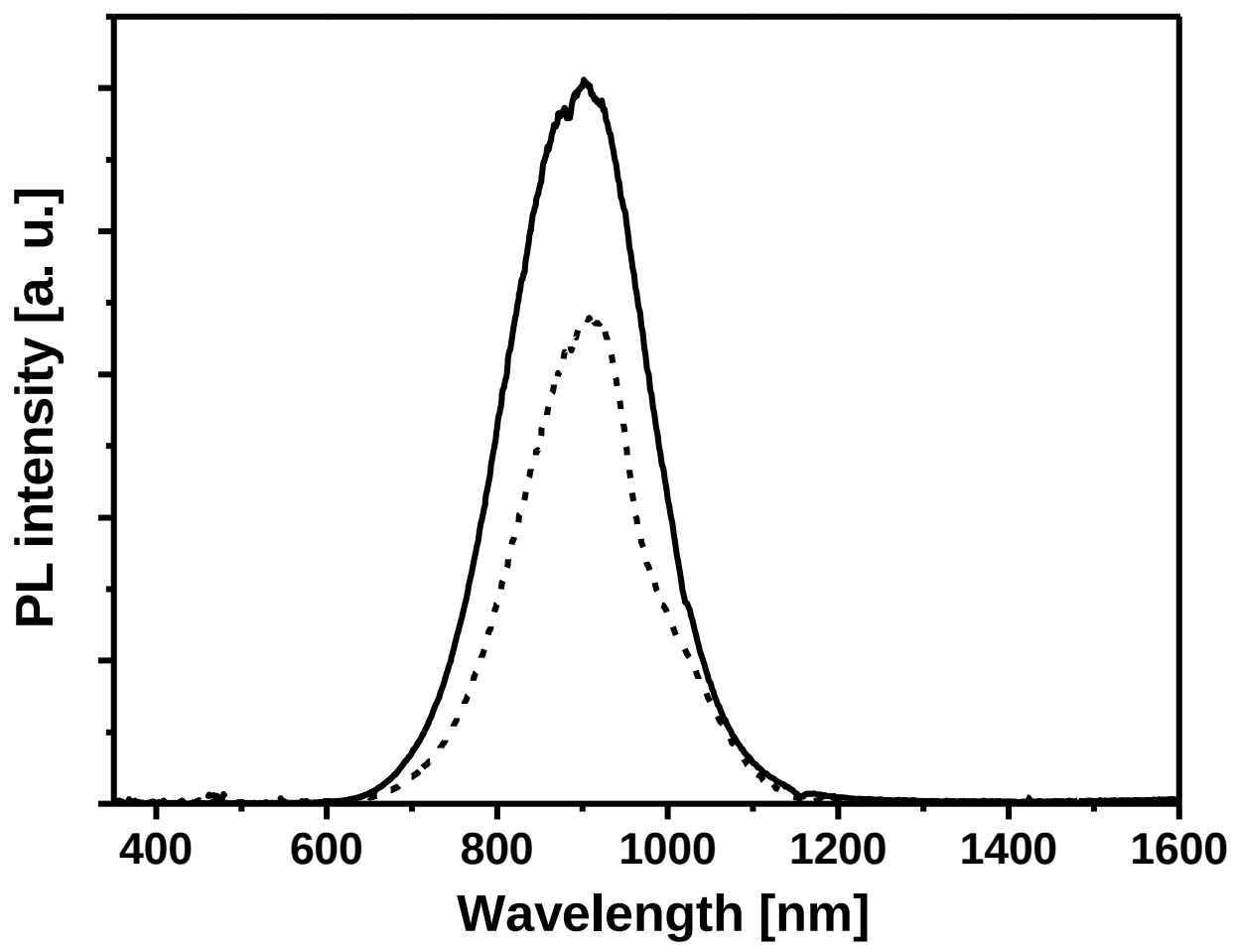

**Figure 11.** PL of as prepared 1.9 nm CSS's in chloroform (solid line) and after transferring them into water (dotted line). An MPA ligand exchange was preformed in order to permit the phase transfer from chloroform solution to water. The QY decreased from 70% in chloroform to 37% in water.

## CONCLUSIONS

Core/shell1/shell2 structures for III-V semiconductor core nanocrystals were developed. Using a layer-by-layer growth technique we can control the composition and structure of core-heteroshell nanocrystals with InAs cores and II-VI semiconductor shells in an unprecedented way. The use of the CdSe buffer layer accompanied by the layer-by-layer growth approach has allowed for a significant improvement in QY over the previously used single shell core/shell particles. We also succeeded in growth of a thick outer shell that provides improved stability and good QY results in water. The new InAs/CdSe/ZnSe CSS structures provide bright tunable emission covering the NIR from below 800 nm to over 1600 nm. These nanocrystals have a narrow distribution of sizes leading also to relatively narrow emission spectra. They also provide exceptional stability and can be implemented in a variety of applications in biological fluorescence tagging and electrooptical devices such as those required in telecommunications fiber-optics.

## Acknowledgment

We Thank Prof. Nir Tesler and Alexey Razin from the Electrical Engineering Faulty in the Technion for help in the quantum-yield measurements. We thank Avi Willenz from the Electron Microscopy Lab of the Life Science Institute for assistance in TEM measurements. We thank Dr. Nira Shimony-Ayal from the Unit for Nanocharacterization for XPS measurements. This work was funded in part by the DIP German-Israel cooperation and the Israel Ministry of Science Tashtiot program.